\documentclass{article}
\usepackage{fortschritte}
\def\beq{\begin{equation}}                     %
\def\eeq{\end{equation}}                       %
\def\bea{\begin{eqnarray}}                     
\def\eea{\end{eqnarray}}                       
\def\cF{{\cal F}} 
\begin {document}  
\begin{flushright} \vspace{-2cm} 
{\small hep-th/0412298} \end{flushright}

\def\titleline{
%
%
%
Topological B-Model and $\hat{c}=1$ Fermionic Strings                       
}
\def\email_speaker{
{\tt 
%
%
}}
\def\authors{
%
%
%
%
%
Yaron Oz                           
}
\def\addresses{
Raymond and Beverly Sackler Faculty of Exact Sciences\\
School of Physics and Astronomy\\
Tel-Aviv University , Ramat-Aviv 69978, Israel
}
\def\abstracttext{
%
%
We construct topological B-model descriptions of 
$\hat{c}=1$ strings, and 
corresponding  Dijkgraaf-Vafa type matrix models
and quiver gauge theories.
}
\large
\makefront
\section{Introduction}

In this lecture  we describe
the relations between four types of systems: non-critical $\hat{c}=1$
 fermionic strings, 
topological strings (B-models), matrix models and ${\cal N}=1$ supersymmetric gauge theories in 
four dimensions  \cite{Ita:2004yn}.
The canonical model that exhibits these relations is 
the non-critical $c=1$ bosonic string at the self-dual radius. 
It has been argued to be equivalent
to the topological B-model on the deformed conifold, and 
the corresponding matrix model and gauge theory are described by the $\hat{A}_1$ quiver diagram
\cite{Dijkgraaf:2003xk,Aganagic:2003qj,Ghoshal:1995wm}.
Since the  $c=1$ bosonic string is well defined only perturbatively, the above connections are 
perturbative.
It is natural to look  
for examples, where the non-critical string is
well defined non-perturbatively.
Natural candidates are the fermionic
$\hat{c}=1$ strings (see \cite{Takayanagi:2003sm,Douglas:2003up} for a matrix model discussion).

A basic ingredient in establishing the connections between these different systems is a 
commutative and associative
ring structure of 
the BRST invariant operators
with zero dimension and zero ghost number.
Its defining relation,
for the $c=1$ bosonic string, 
is the conifold equation \cite{Witten:1991zd}.
For the fermionic
$\hat{c}=1$ strings with a non-chiral GSO projection, the ring structure
corresponds to certain $Z_2$ quotients of the conifold \cite{Ita:2004yn}, while with a 
chiral GSO projection one gets a Calabi-Yau 3-fold suspended pinched point
singularity \cite{us}.

The partition functions of the non-critical strings 
are matched with the partition functions of the topological B-model on the (deformed) 
Calabi-Yau singularities, with the partition functions of the 
quiver matrix models and with 
 the glueball F-terms of four-dimensional 
gauge theories.

\section{$\hat{c} = 1$ Fermionic Strings}

Fermionic strings are described by $N=1$ supersymmetric worldsheet 
field theories coupled to worldsheet supergravity.
In the superconformal
gauge it is described by a matter superfield $X$ and the super
Liouville field $\Phi$. 
In components they take the form
\bea
X &=& x+\theta\psi_x+\bar\theta\bar\psi_x+\theta\bar\theta F_x \ , \nonumber\\
\Phi_l &=&\phi_l+\theta\psi_l+\bar\theta\bar\psi_l+\theta\bar\theta F_l \ .
\eea
The field $X$ is free, while $\Phi$ is described by the super Liouville
Largangian.

We consider the theory compactified on a circle of radius $R$.
One can perform a chiral GSO projection giving Type II strings, and  
a nonchiral GSO projection which gives type 0
string theory.
In this lecture we will consider the latter case.
There are two distinct choices, depending on how the
worldsheet fermion number $(-1)^F$ symmetry is realized in the closed string 
R-R sector, they are called (circle line) $0$A and $0$B.
In both theories there 
are no $(NS,R)$ or $(R,NS)$ sectors, and therefore no spacetime fermions.
The `super affine' theories are obtained from them 
by modding out by a $Z_2$ symmetry $(-)^{{\bf F}_L} e^{i\pi p}$
where $(-)^{{\bf F}_L}=1\;(-1)$ on NS-NS (R-R) momentum states, and $p$ 
is the momentum $k=\frac{p}{R}$.
Type 0A and Type 0B are T-dual
to each other under
$R\rightarrow{\alpha'\over R}$.
The super affine theories are
self dual under
$R\rightarrow{2\alpha'\over R}$.

We will consider the theories at the special radii
\beq
R_{\rm 0A}= \frac{l_s}{\sqrt{2}},~~~~~R_{\rm 0B}= l_s\sqrt{2},~~~~
R_{\rm super-affine}=l_s\sqrt{2} \ ,
\label{topp}
\eeq
and
argue that at these points the non-critical $\hat{c}=1$ string theories have
a description as a topological B-model on a Calabi-Yau 3-fold.
At these radii the winding modes contribute exactly as the momentum 
modes to the torus partition function, and we get the results of table 1.
\begin{table}
\hspace{2cm}
\begin{tabular}{||l|l||l|l||}\hline $F_{\rm c=1}(\rm R_{self-dual})
$&$-\frac{1}{12}\ln\mu$&&\\\hline
$F_{\rm 0A}(\sqrt{\alpha'/2})$&$-\frac{1}{6}\ln\mu$&$F_{\rm 0B}
(2\sqrt{\alpha'/2})$&$-\frac{1}{6}\ln\mu$\\\hline
$F_{\rm 0A}^{\rm super-aff.}(R_{\rm self-dual})$&$-\frac{1}{6}\ln\mu$&&\\\hline
$F_{\rm 0B}^{\rm super-aff.}(R_{\rm self-dual})$&$-\frac{1}{12}\ln\mu$&&\\\hline
\end{tabular}\\\\
\caption{Torus partition functions at the special radii}
\label{table1}
\end{table}

In the following we will use the convention $\alpha^{\prime}=2$.

\section{The Ground Ring and Topological B-model}
The
spin zero ghost number zero BRST invariant operators generate a commutative,
 associative
ring
\beq
{\cal O}(z){\cal O}'(0)\sim {\cal O}''(0)+\{Q,\dots\} \ ,
\label{BRS}
\eeq
called the ground ring, where $Q$ is the BRST operator.

The chiral BRST cohomology of dimension zero and ghost number zero
 is given by the 
infinite set of states
$\Psi_{(r,s)}$ with $r,s$ negative integers \cite{Bouwknegt:1991am,Bouwknegt:1991va,Itoh:1991ix}
\beq
\Psi_{(r,s)}\sim O_{r,s} e^{\left(ik_{r,s} x_L-p_{r,s}\Phi_L\right)} \ .
\eeq
The Liouville and matter momentum are given by 
\beq
k_{r,s}=\frac{1}{2}(r-s),\quad p_{r,s}=\frac{1}{2}(r+s+2) \ .
\eeq
The operators $\Psi_{(r,s)}$  are in the NS-sector if 
$k_{r,s}=(r-s)/2$ takes integer values, and in the R-sector if it takes half integer values. 

The basic elements for the construction of the ring are the
R-sector operators
\beq
x(z)\equiv \Psi_{(-1,-2)}(z),~~~  
y(z)\equiv \Psi_{(-2,-1)}(z) \ , 
\eeq
and the NS-sector operators 
\beq
\label{NSCP1}
u(z)\equiv\Psi_{(-1,-3)}(z)=x^2,~~~ 
v(z)\equiv\Psi_{(-3,-1)}(z)=y^2,~~~ 
w(z)\equiv\Psi_{(-2,-2)}(z)=xy \ .
\eeq

In order to construct the ground ring, we combine the left and right sectors
with the same left and right Liouville momenta.
Denote:
\begin{eqnarray}\label{ringelements}
  \begin{array}{ccc}
a_{ij}=\left(\begin{array}{cc}x\bar x&x\bar y\\
                         y\bar x&y\bar y \end{array}\right),
&\quad\quad&b_{ij}=\left(\begin{array}{ccc} u\bar u & u\bar w & u\bar v\\
                           w\bar u & w\bar w & w\bar v\\
                           v\bar u & v\bar w & v\bar v\end{array}\right) \ .
\end{array}
\end{eqnarray}
Note that 
\beq
det(a_{ij}) = 0 \ ,
\label{det}
\eeq
which is the conifold equation.

\vskip 0.1cm
\noindent
{\it The ground rings}\\

Imposing the GSO projection we get:

\begin{itemize}

\item{} Circle line theories:
The ring is
generated by four elements   $a_{12}, a_{21}, b_{11},b_{33}$
with the relation
 \begin{eqnarray}
\label{GR0AB1} 
 (a_{12})^2(a_{21})^2-b_{11}b_{33}=0 \ .
\label{conz2}
\eea
The complex 3-fold (\ref{conz2}) is the $Z_2$ quotient of the conifold
(\ref{det}), with the $Z_2$ action being
\bea
Z_2:~~~~(a_{11},a_{22}) &\rightarrow& -(a_{11},a_{22}) \nonumber\\
(a_{12},a_{21}) &\rightarrow& (a_{12},a_{21}) \ .
\label{Z2}
\end{eqnarray}

When the cosmological constant $\mu$ and the background RR charge $q$
are nonzero we get a deformation of the ground ring 
relation
\beq
\label{Geomy11}
(a_{12}a_{21} +\mu)^2= b_{11}b_{33} -\frac{q^2}{4} \ .
\end{equation}

\item{} Super affine 0A: 
The ground ring is generated by
the invariant elements $b_{ij}$ subject to the conditions (\ref{det}) and the projection 
 $a_{ij}\rightarrow -a_{ij}$. 
The complex 3-fold described by the ground ring  is the $Z_2$ quotient of the conifold
(\ref{det})
\beq
Z_2:~~~~a_{ij} \rightarrow -a_{ij} \ .
\label{z22}
\eeq
The singular geometry described by the ground ring
is that of a Calabi-Yau space, where a three cycle of the form $S^3/Z_2$
shrinks to zero size.
When $\mu\neq0$ the deformed space is such that
$Vol(S^3/Z_2) \sim \mu/2$, and is locally $T^*(S^3/Z_2)$.

\item{} Super affine 0B: 
The ground ring is generated by
the invariant elements $a_{ij}$ subject to the conditions (\ref{det}).
The complex 3-fold described by the ground ring  is the conifold.
When $\mu \neq 0$ the ground ring is described by
the deformed conifold.

\end{itemize}

\noindent
{\it Topological B-models}\\

Based on the ground rings structures we predict the following
dualities: 

\begin{itemize}

\item{} Type 0A (0B) at the radius $R=1$ ($R=2$) is equivalent to the 
topological B-model on the deformed $Z_2$ quotient of the
conifold (\ref{Z2}).

\item{} Super affine  0A at the radius $R=2$ is equivalent to the
topological B-model on the deformed $Z_2$ quotient of the
conifold (\ref{z22}).

\item{} Super affine  0B at the radius $R=2$ is equivalent to  the
topological B-model on the deformed
conifold.

\end{itemize}

Note that since the $\hat{c}=1$ non-critical strings are well defined non-perturbatively,
the above dualities provide a non-perturbative definition of the topological
strings.

\vskip 0.1cm
\noindent
{\it Partition functions}\\

We find at the special radii:

\begin{itemize}

\item{}Circle line theories: 

\begin{equation}
\label{C1vs0A}
{\cal F}_{\rm 0A}(R=1)={\cal F}_{c=1}\left(\mu+\frac{iq}{2}\right)
+{\cal F}_{c=1}\left(\mu-\frac{iq}{2}\right) \ ,
\end{equation}
where ${\cal F}_{c=1}$ is the partition function
of the $c=1$ bosonic string 
at the self-dual radius, or equivalently 
the partition function of the topological B-model on the deformed
conifold.
The relation (\ref{C1vs0A}) is perturbative.
When $q=0$ we have
\beq
{\cal F}_{\rm 0A}(R=1)=
2\cF_{c=1}(R_{\rm self-dual}) \ ,
\label{part2}
\eeq
where
\bea
\cF_{c=1}(R_{\rm self-dual}) ={1\over 2}\mu^2 \log\mu -{1\over 12} \log\mu +
{1\over 240}\mu^{-2}+\sum_{g>2} a_g \mu^{2-2g} \ .
\label{part}
\eea
$a_g = \frac{B_{2g}}{2g(2g-2)}$ is the Euler class of 
the moduli space of Riemann surfaces of genus $g$. 

\item{}Super affine 0A: 

We find that {\it non-perturbatively} at the special radii (\ref{topp})
\beq
\cF^{\rm super-aff.}_{\rm 0A}(R=1) = \cF_{\rm 0A}(R=2) \ ,
\eeq
when the RR charge $q=0$.

\item{}Super affine 0B:

We suggest that while perturbatively
\bea
\cF^{\rm super-aff}_{\rm 0B}(R=2)=
\cF_{c=1}(R_{\rm self-dual}) \ ,
\label{part1}
\eea
it provide a {\it non-perturbative} 
completion of topological B-model on the deformed
conifold.

\end{itemize}

\section{Matrix Models and Quiver Gauge theories}

Consider D-branes wrapping holomorphic cycles in Calabi-Yau
3-folds. 
In the language of four-dimensional ${\cal N}=1$ supersymmetric gauge theory,
the D-branes wrapping in the resolved geometry provide the UV description,
while the IR physics is described by the deformed geometry after the transition.
For confining gauge theories, one assumes that the relevant IR degrees of freedom
are the glueball superfields $S_i$. 
The partition function of the topological field theory, as a function of the deformation 
parameters, computes the holomorphic F-terms of the gauge theory as a function
of the glueball superfields.
This has also a matrix model description 
\cite{Dijkgraaf:2002fc,Dijkgraaf:2002vw,Dijkgraaf:2002dh,Dijkgraaf:2003xk}.

Consider as an example
the super affine 0A theory at the self-dual radius.
In order to construct the quiver gauge theory, we work in 
the semi classical regime, where the resolved geometry is described by the $Z_2$
quotient of the resolved conifold, and perform
a $Z_2$ quotient of a system
of $N_0$ D3-branes 
and $2N_1$ D5-branes wrapping the resolved conifold.
The gauge and matter content of the resulting quiver theory
reads
\begin{eqnarray}
  \begin{array}{ccccc}&SU(N)&SU(N)&SU(K)&SU(K)\\
    A_i&N&&\bar K&\\
    \tilde{A}_i&&N&&\bar K\\
    B_j&\bar N&&&K\\
    \tilde{B}_j&&\bar N&K&\\
    \Phi^+_1&N&\bar N&&\\
    \Phi^+_2&\bar N&N&&\\
    \Phi^-_1&&&K&\bar K\\
    \Phi^-_2&&&\bar K&K\end{array}
\end{eqnarray}
with $i=1,2$, and $N=N_0, K=N_0+N_1$.
The superpotential reads 
\begin{eqnarray}
  W_{tree}&=&W_{0}+W_{1}\nonumber\\
  W_{0}&=&m\, tr(\Phi^+_1\Phi^+_2)-m\, tr(\Phi^-_1\Phi^-_2) \ , \nonumber\\
  W_{1}&=&-tr(\tilde A_i\Phi^+_1\tilde B_i)-tr(A_i\Phi^+_2B_i)-
tr(\tilde B_i\Phi^-_1 A_i)-tr(B_i\Phi^-_2\tilde A_i) \ .
\label{sup}
\end{eqnarray}

Consider next the IR dynamics of the 
quiver gauge theory. 
Of particular interest for us is the case
of $N_0$ being an integer multiple of $N_1$.
In this case, after the duality cascades, one ends up with a
confining gauge theory with 
the gauge group $SU(N_1)^2$,
and the massive bifundamentals
$\Phi_i^+, i=1,2$.
In the deformed geometry $N_1$ 
sets the size of $S^3$ which we identify with the
glueball superfield $S$.
The holomorphic F-terms of this theory ${\cal F}(S)$ as a function of  $S$ are twice that
of  $SU(N_1)$ SYM and are related to the perturbative super affine 0A free energy
as
\beq
{\cal F}_{\rm SYM}(S) = \cF_{\rm 0A}^{\rm super-aff.}(R_{\rm self-dual})(\mu) \ ,
\eeq
where $S\sim \mu$.

One can also write a 
DV matrix integral description of the quiver gauge theory F-terms.
It takes the form:
\beq
Z = {1\over V}\int d\Phi_i^+ d\Phi_i^- dA_i dB_j d\tilde{A}_i d\tilde{B}_j exp
\left[-{1\over g_s} W_{tree}(\Phi_i^+, \Phi_i^- A_i,B_j, \tilde{A}_i, \tilde{B}_j)\right] \ ,
\eeq
with  $W_{tree}$ given by (\ref{sup}), and $V$ is the volume of the groups
$SU(\hat{N_0})^2\times SU(\hat{N_0}+\hat{N_1})^2$.
Similar cascade arguments lead
to the matrix integral 
\begin{eqnarray}
 Z=\frac{1}{V(SU(\hat{N})\times SU(\hat{N}))}\int d\Phi^{+}_{1}d\Phi^{+}_{2} e^{-\frac{m}{g_s}\,tr(\Phi^+_1\Phi^+_2)}
\ ,
\end{eqnarray}
which gives perturbatively
\beq
 {\cal F}_{\rm matrix~ model}(S) = 2\,\times\,{\cal F}_{c=1}(R_{\rm self-dual})(\mu) 
= {\cal F}_{\rm 0A}^{\rm super-affine}(R_{\rm self-dual})(\mu) \ ,
\eeq
with $S=g_s \hat{N}$.

{\bf Acknowledgement} I would like to thank H. Ita, H. Nieder
and T. Sakai for collaboration on the work presented here, and for many
valuable discussions.

\end{document}